\documentclass[final,11pt,2p,times,sort&compress]{elsarticle} 

  \usepackage{color}
  \usepackage{amsmath}
  \usepackage{calligra}
  \usepackage{caption}
  \usepackage{cprotect}
  \usepackage{enumerate}
  \usepackage[T1]{fontenc}
  \usepackage{ulem}
  \usepackage{lineno}
  \usepackage{multicol}
  \usepackage{tabularx}
  \usepackage[]{titlesec}
  \newcolumntype{L}[1]{>{\raggedright\arraybackslash}p{#1} }
  \newcolumntype{C}[1]{>{\centering  \arraybackslash}p{#1} }
  \newcolumntype{R}[1]{>{\raggedleft \arraybackslash}p{#1} }

\usepackage{mhchem}
\usepackage{longtable} 
\usepackage{pdflscape} 

\journal{Scripta Materialia}

\begin{document}

\begin{frontmatter}
  \title{High Throughput combinatorial method for fast and \\ robust prediction of lattice thermal conductivity}
  \author{Pinku Nath$^1$, Jose J. Plata$^1$, Demet Usanmaz$^1$, Cormac Toher$^1$,\\Marco Fornari$^2$, Marco Buongiorno Nardelli$^3$, Stefano Curtarolo$^{4,\star}$}
  \address{$^{1}$ Department of Mechanical Engineering and Materials Science, Duke University, Durham, North Carolina 27708, USA.}
  \address{$^{2}$ Department of Physics and Science of Advanced Materials Program, Central Michigan University, Mount Pleasant, MI 48858, USA.}
  \address{$^{3}$ Department of Physics and Department of Chemistry, University of North Texas, Denton TX, USA.}
  \address{$^{4}$ Materials Science, Electrical Engineering, Physics and Chemistry, Duke University, Durham NC, 27708, USA.}
  \address{$^{\star}${\bf corresponding:} stefano@duke.edu}
  \begin{abstract}
The lack of computationally inexpensive and accurate ab-initio based methodologies to predict lattice thermal conductivity, without
 computing the anharmonic force constants or time-consuming ab-initio molecular dynamics, is one of the obstacles preventing the accelerated discovery
of new high or low thermal conductivity materials. 
The Slack equation is the best alternative to other more expensive methodologies but is highly dependent on two variables: the acoustic Debye temperature, $\theta_a$, and the Gr\"{u}neisen parameter, $\gamma$. 
Furthermore, different definitions can be used for these two quantities depending on the model or approximation. 
In this article, we present a combinatorial approach to elucidate which definitions of both variables produce the best predictions of the lattice thermal conductivity, $\kappa_l$.
A set of 42 compounds was used to test accuracy and robustness of all possible combinations. 
This approach is ideal for obtaining more accurate values than fast screening models based on the Debye model, while being significantly less expensive than methodologies that solve the Boltzmann transport equation.
  \end{abstract}
\end{frontmatter}

\section{Introduction}

Lattice thermal conductivity, $\kappa_l$, plays an important role in  multiple applications and 
technologies \cite{curtarolo:art81}. 
Thermoelectric materials~\cite{zebarjadi_perspectives_2012}, heat sink materials 
\cite{Yeh_2002}, rewritable density scanning-probe phase-change memories \cite{Wright_tnano_2011} or 
thermal medical therapies \cite{Cahill_apr_2014} are some examples in
which thermal transport is the technological enabling property.
During the last three  decades, several theoretical models and
methodologies have been developed to calculate $\kappa_l$ 
\cite{Ziman_1960,callaway_model_1959,Allen_PHMB_1994,Broido2007,Wang2011}. 
It is a trade-off: while the most quick approaches can only predict
trends, accurate methods are computationally expensive and can not be
implemented in high throughput (HT) frameworks, hindering the
discovery of new materials with better performance \cite{curtarolo:art81,curtarolo:art84,curtarolo:art85}.
For instance, semi-empirical models predict thermal properties at reduced computational cost but require some 
experimental data \cite{Ziman_1960,callaway_model_1959,Allen_PHMB_1994}.
Additionally, classical molecular dynamics combined with Green-Kubo
equations also produces reliable results. Although this method includes 
high-order scattering processes, the us of semi-empirical potentials
leads to errors on the order of 50\%
\cite{Green_JCP_1954,Kubo_JPSJ_1957}.
Furthermore, the extension of the Green-Kubo formalism to multiscale
models has been shown to be non trivial \cite{curtarolo:art12}.
Characterization of the anharmonic forces constants and its use in the solution of the Boltzmann transport equation, BTE, is extremely expensive which makes it unfeasible in a HT approach. 
Thus, for an effective prediction of $\kappa_l$ with HT methods,
there is an advantate in approximated methods based on \textit{ab-initio} characterization. 
{\bf i.} {\it Ab-initio} is fully self-consistent and does not need experimental data or the use/generation of force fields;
{\bf ii.} There exists HT frameworks, such as {\small{AFLOW}} \cite{aflowPAPER,aflowlibPAPER,Calderon_cms_2015}, that can monitor, manage, correct, and post-process the information obtained from different quantum mechanical codes.

There are three main families of approximated models based on first
principles. They can be classified depending on performance, accuracy and robustness.

\begin{itemize}

\item The ``GIBBS'' quasiharmonic Debye model is the least computationally expensive approach to identify trends and simple descriptors for thermal properties, such as the Gr\"{u}neisen parameters and the Debye temperature \cite{Blanco_CPC_GIBBS_2004}. 
The \underline{A}flow-\underline{G}ibbs-\underline{L}ibrary
  (AGL) framework  combines this model with the Slack equation \cite{slack}  based on work with noble gas crystals by Leibfried and Schlomann \cite{Leibfried_1965}  and Julian \cite{Julian_PRev_1965}  to predict $\kappa_l$ \cite{curtarolo:art96}. 
It reproduces correctly the ordinal ranking of the thermal conductivity for several different classes of semiconductor materials using only energy-volume curves, but suffers in comparing families with different structures.

\item More accurate Gr\"{u}neisen parameters can be obtained using the quasiharmonic approximation, QHA,  and then used in the Slack equation \cite{Madsen_PRB_2014}.
However, harmonic force constants have to be calculated to build the dynamical matrix that describe the vibrational modes of the system.  
An alternative method proposed by Madsen \textit{et al}. \cite{Madsen_PRB_2014} consists of the use of lattice dynamics calculations to compute approximate  relaxation scattering times at $\theta_D$. Both methods give $\kappa_l$ in
reasonable quantitative agreement with experiments.

\item Third order interatomic force constants (3rd IFCs)  are required to calculate the phonon scattering times included in the  solution of the Boltzmann transport equation, BTE \cite{Broido2007,Wang2011,Deinzer_PRB_2003}.
This is the most computationally expensive method but it is the most used for highly accurate results. 
Once scattering processes are computed using 3rd IFCs, different schemes have been proposed to solve the BTE. 
The relaxation time approximation, RTA, is the simplest and predicts values on average 10\% smaller than experimental quantities. 
The full  solution requires a self-consistent iteration, but produces  values very close to experiment while adding  only
 a small computational cost compared to RTA solutions \cite{Mingo_book_2014}.
The bottle neck of both methods comes from  on the computation of the 3rd order IFCs. 
Recently, some authors have proposed the computation of these forces using compressive sensing. However, 
the cost to obtain reliable results is still high \cite{Fei_PRL_2014}. 

\end{itemize}

Despite  the different approaches developed in the last few decades, there is a lack of inexpensive, accurate, and robust  methods that can be used routinely. 
In this article, we use different definitions of $\theta_a$ and $\gamma$  based on the Debye model and the QHA to obtain values for $\kappa_l$ using the Slack equation.
Using a phenomenological approach, we compare the results obtained with our combinatorial schema to available experimental data to decide which description of these two variables best predict values for $\kappa_l$ with qualitative and quantitative accuracy in a high-throughput approach.  

\section{Methodology }

\subsection{Lattice thermal conductivity}

The lattice thermal conductivity is computed using the Slack equation (Eq. \ref{slack}) because of its simplicity: 
\begin{eqnarray}
 \label{slack}
 \kappa_l (\theta_a) &=& \frac{0.849 \times 3 \sqrt[3](4)}{20 \pi^3(1 - 0.514\gamma^{-1} + 0.228\gamma^{-2})} \times \\ \nonumber
 & & \times \left( \frac{k_B \theta_a}{\hbar} \right)^2 \frac{k_B M_{\mathrm{av}} V^{\frac{1}{3}}}{\hbar \gamma^2}.
\end{eqnarray}
This equation predicts $\kappa_l$ at the acoustic Debye temperature ($\theta_a$) using  the Gr{\"u}neisen parameter, $\gamma$, 
primitive cell volume, $V$,  and the average atomic mass, $M_{\mathrm{av}}$. 
The thermal conductivity at any temperature is estimated by \cite{slack, Morelli_Slack_2006, Madsen_PRB_2014}:
\begin{equation}
 \label{kappa_T}
\kappa_l(T) = \kappa_l(\theta_a) \frac{\theta_a}{T}.
\end{equation}
Equations (\ref{slack}) and (\ref{kappa_T}) show that only the acoustic Debye temperature and the  
Gr{\"u}neisen parameter are needed to calculate $\kappa_l$ at any temperature. 
Various groups have used different models to predict this quantity using the Slack
equation.
%
%
It seems that Slack equation combined with the Debye model tends to underestimate $\kappa_l$ \cite{curtarolo:art96}, while when 
combined with lattice dynamics calculations, $\kappa_l$ is overestimated \cite{Madsen_PRB_2014}. 
We propose the combination of both models to offset the errors and obtain values closer to the 
experimental results. 

\begin{figure}[]
 \includegraphics[width=0.98\columnwidth]{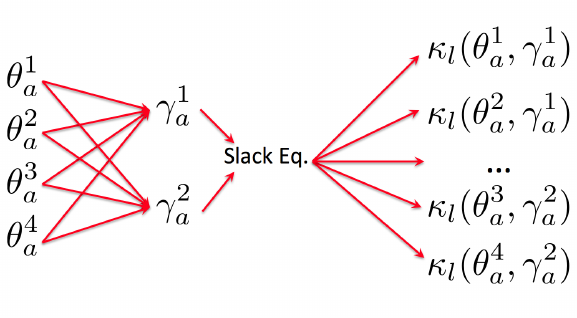}
 \vspace{-4mm}
 \caption{ Combinatorial diagram for  lattice thermal conductivity}
 \label{fig:schema}
\end{figure}

Our implementation is based on a combinatorial approach where the lattice thermal conductivity is a function of two variables: acoustic Debye 
temperature, $\theta_a$,  and Gr{\"u}neisen parameter, $\gamma$:  $\kappa_l(\theta_a,\gamma)$.
We use different formulations based on the Debye model and quasiharmonic approximation to compute these two 
variables and then combine them to obtain 
different values for $\kappa_l(\theta^x_a,\gamma^y)$ (see Figure \ref{fig:schema}).
We use a set of 42 materials, all  belonging to different space groups and presenting  a range of four orders of magnitude for
$\kappa_l$, as a test to determine the best combination of variables from a quantitative point of view.
This approach maximizes the flexibility of the method, optimizing the 
results without extra costs beyond the harmonic force constants calculations.  

The first definition for $\gamma$ is extracted from Ref. \cite{Madsen_PRB_2014}:
\begin{equation}
 \label{gru_madsen}
\gamma^{(1)}_a=\sqrt{\overline{\gamma}^2_a}=\sqrt{\frac{\sum\limits_{q}\sum\limits_{i}^{\hbar\omega_{iq}<k_B\theta_a}\gamma^{2}_{iq} C_{iq}}{\sum\limits_{q}\sum\limits_{i}^{\hbar\omega_{iq}<k_B\theta_a}C_{iq}}},
\end{equation}
where $\gamma_{iq}$ is defines as:
\begin{equation}\label{EQ:GAMMAQJD}
 \gamma_{iq}=-\frac{V_{eq}^{0K}}{2\nu_{qj}^2}\sum_i e_{iq} \frac{\partial D_q}{\partial V} e_{iq}^*.
 \end{equation}
Since the acoustic bands provide the majority of the contribution to $\kappa_l$,  the sum is performed over 
the modes, $i$, and q-points, $q$, having an energy less than $k_B\theta_a$.
We propose a second definition where the sum is done only for the acoustic modes instead of over the 
modes having an energy less than $k_B\theta_a$: 
\begin{equation}
 \label{gru_ac}
\gamma^{(2)}_a=\sqrt{\overline{\gamma}^2_a}=\sqrt{\frac{\sum\limits_{q}\sum\limits_{i=1}^{3}\gamma^{2}_{a,iq} C_{iq}}{\sum\limits_{q}\sum\limits_{i=1}^{3}C_{iq}}}.
\end{equation}
It is important to note that while $\gamma^{(1)}_a$ depends on $\theta_a$,  $\gamma^{(2)}_a$ is 
independent of the other variable.

Recently, Madsen \textit{et al.} also used the averaged squared Gr\"{u}neisen parameters obtaining similar results with their model \cite{Madsen_pssa_2016}:
\begin{equation}
 \label{gru}
\gamma^{(3)}=\sqrt{\overline{\gamma}^2}=\sqrt{\frac{\sum\limits_{q}\sum\limits_{i}\gamma^{2}_{iq} C_{iq}}{\sum\limits_{q}\sum\limits_{i}C_{iq}}}.
\end{equation}

Phonon calculations can be also used to predict $\theta_a$:
\begin{equation}
 \label{theta_1}
\theta^{(1)}_a = n^{1/3} \theta_D = n^{1/3} \sqrt{\frac{5 \hbar}{3k_{B}}\frac{\int_{0}^{\infty} \omega^2g(\omega)d\omega}{\int_{0}^{\infty} g(\omega)d\omega}}.
\end{equation}
where $n$ is the number of atoms per unit cell and $g(\omega)$ is the phonon density of states. 
Similar to  $\gamma^{(2)}_a$, we  get $\theta_a$ using only the  acoustic branches:
\begin{equation}
 \label{theta_2}
\theta^{(2)}_a = \sqrt{\frac{5 \hbar}{3k_{B}}\frac{\int_{0}^{\infty} \omega_a^2g(\omega_a)d\omega_a}{\int_{0}^{\infty} g(\omega_a)d\omega_a}}.
\end{equation}
The classic definition for the Debye frequency, the maximun frequency in the vibrational spectra,  can be also applied to obtain the Debye temperature using the phonon density of states, pDOS:
\begin{equation}
 \label{debye_frec}
\int^{\omega_D}_0g(\omega)d\omega=3n,
\end{equation}
\begin{equation}
 \label{theta_3}
 \theta^{(3)}_a=n^{1/3} \theta_D=n^{1/3} \frac{\hbar\omega_D}{k_B}.
\end{equation}

The Debye model also predicts $\theta_D$ using the bulk modulus, $B$ \cite{Blanco_CPC_GIBBS_2004,Blanco_jmolstrthch_1996,Poirier_Earth_Interior_2000}:
\begin{equation}
 \label{theta_4}
 \theta^{(4)}_a=n^{1/3} \theta_D = n^{1/3} \frac{\hbar}{k_B}[6 \pi^2 V^{1/2} n]^{1/3} f(\sigma) \sqrt{\frac{B}{M}}.
\end{equation}
Here, $M$ is the mass of the unit cell,  and $f(\sigma)$ is given by
\begin{equation}
 \label{fpoisson}
 f(\sigma) = \left\{ 3 \left[ 2 \left( \frac{2}{3} \cdot \frac{1 + \sigma}{1 - 2 \sigma} \right)^{3/2} + \left( \frac{1}{3} \cdot \frac{1 + \sigma}{1 - \sigma} \right)^{3/2} \right]^{-1} \right\}^{\frac{1}{3}},
\end{equation}
within the assumption that the Poisson ratio, $\sigma$, remains constant.
For the calculations described in this article, this value is set at 0.25, which is the theoretical value for a Cauchy solid \cite{Blanco_CPC_GIBBS_2004, Poirier_Earth_Interior_2000}. 
The Poisson ratio $\sigma$ for crystalline materials is typically in the range 0.2-0.3. 
Bulk modulus, $B$, is obtained fitting free energy, $F(V, T)$, obtained using the QHA to the Birch-Murnaghan (BM) function:
\begin{equation}\label{EQ:BM}
F(V) = F_{eq} + \frac{BV_{eq}}{B_p}\Bigg[ \frac{(V_{eq}/V)^{B_p}}{Bp-1} +1\Bigg] - \frac{V_{eq}B}{Bp-1}, 
\end{equation}
 where, equilibrium free energy, $F_{eq}$, bulk modulus, $B$, equilibrium volume, $V_{eq}$ and the derivative of the bulk modulus with respect to pressure, $B_p$ are used as the fitting parameters.

\subsection{Computational details}

\subsubsection*{Geometry optimization}

All structures are fully relaxed  using the automated framework, {\small AFLOW} \cite{aflowPAPER,aflowlibPAPER,Calderon_cms_2015},
and the DFT Vienna Ab-initio simulation package, {\small VASP} \cite{kresse_vasp}. Optimizations are performed following the
{\small AFLOW} standards \cite{Calderon_cms_2015}. We use the projector augmented wave (PAW) pseudopotentials \cite{PAW}
and the exchange and correlation functionals parametrized by the generalized gradient approximation proposed
by Perdew-Burke-Ernzerhof (PBE) \cite{PBE}. All calculations use a high energy-cutoff, which is 40$\%$ larger
than the maximum recommended cutoff among all component potentials, and a {\bf k}-points mesh of 8,000 {\bf k}-points per reciprocal atom.
Primitive cells are fully relaxed (lattice parameters and ionic positions) until the energy difference between two consecutive ionic steps
is smaller than $10^{-4}$ eV and forces in each atom are below $10^{-3}$ eV/\AA.

\subsubsection{Phonon calculations}

Phonon calculations  were  carried out  using the automatic phonon library, APL, as implemented in the {\small AFLOW}
package, using  VASP  to obtain the IFCs via the finite-displacement approach \cite{Pinku_sr_2016}.
The magnitude of this displacement is 0.015 \AA. 
Electronic self consistent field (SCF) for single point calculations were stopped when the difference of
energy between last two step was less than $10^{-8}$ eV. This threshold ensures a good convergence
for the wavefunction and accurate enough values for forces and harmonic force constants.
Non-analytical contributions to the dynamical matrix are also included using the formulation developed by Wang \textit{et al} \cite{Wang2010}.
Frequencies and other related phonon properties are calculated on a  21$\times$21$\times$21 q-mesh in the Brillouin zone, which is sufficient
to converge the  vibrational  density of  states, pDOS,  and  hence  the  values  of thermodynamic properties calculated through it.
The pDOS is calculated using the linear interpolation tetrahedron method available in {\small AFLOW} package.
The derivative of dynamical matrix is obtained using the central difference method within a 
volume range of $\pm 0.03 \%$.

\section{Results}

A  data set of 42 compounds has been used to validate our approach.
The list of materials includes semiconductors and insulators that belong to different
structure prototypes such as diamond, zinc blende, rock salt and  fluorite.
To maximize the heterogeneity of the data set, materials have been selected containing as many different elements as possible from the s-, p-, and d-blocks of 
the periodic table.

\begin{figure*}
\includegraphics*[width=0.98\columnwidth]{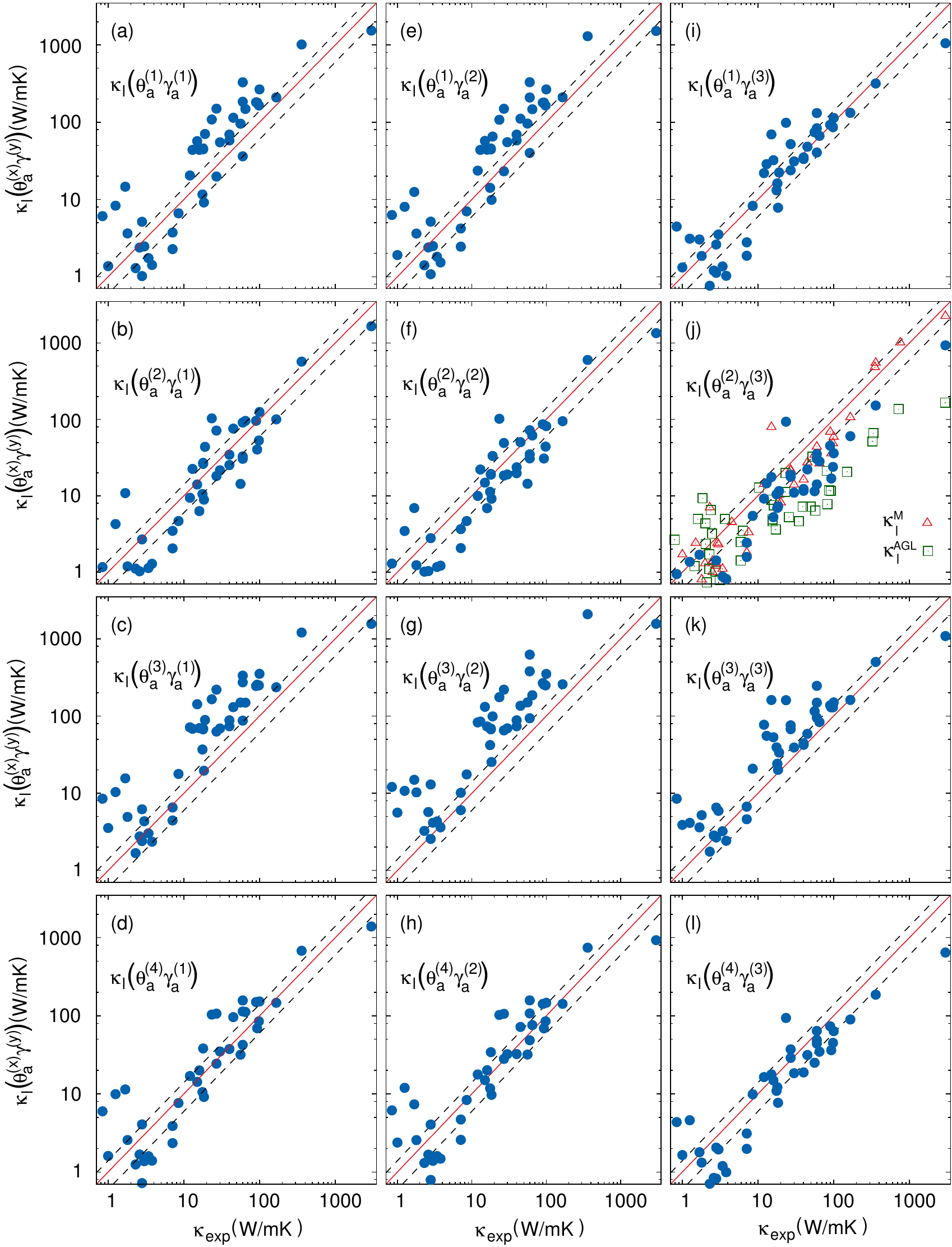}
\caption{ Lattice thermal conductivity, $\kappa_l(\theta_a,\gamma)$, for the proposed models. Dashed black lines represent $\pm$40\% error.}
\label{fig:kappa_panel}
\end{figure*}

The comparison between our combinatorial approach and experimentally reported values for $\kappa_l$ is depicted in 
Figure \ref{fig:kappa_panel}.
%
%
The twelve models predict qualitatively the main trend for $\kappa_l$.
We used different statistical quantities to measure the level of qualitative and 
quantitative agreement between the models and the experimental data.
The Pearson and Spearman correlations measure the linear correlation and the 
monotonicity of the relationship between the two variables respectively 
(see Table \ref{tab:stats}).
We have found high Pearson and Spearman correlation values for some of the predicted models. For instance,
$\kappa_l\left(\theta^{(1)}_a,\gamma^{(3)}\right)$, $\kappa_l\left(\theta^{(2)}_a,\gamma^{(3)}\right)$ and
$\kappa_l\left(\theta^{(4)}_a,\gamma^{(3)}\right)$ present Pearson and  
Spearman correlations with experiments higher than 0.90.

 \begin{table}[h]
     \caption{Pearson and (Spearman) correlations for the material data set. Highest Pearson correlations in bold.}
     \centering
     \begin{tabular}[b]{c C{1.2cm} C{1.2cm} C{1.2cm} C{1.2cm} }
       \hline \hline
                & $\theta^{(1)}_a$ & $\theta^{(2)}_a$ & $\theta^{(3)}_a$ & $\theta^{(4)}_a$  \\
       \hline
        $\gamma^{(1)}_a$  &0.87 (0.88) & 0.97 (0.89)                  & 0.82 (0.90)   & 0.93 (0.88)          \\
        $\gamma^{(2)}_a$  &0.80 (0.89) & 0.95 (0.89)                  & 0.64 (0.89)   & 0.83 (0.89)          \\
        $\gamma^{(3)}$    &0.97 (0.90) & \textbf{0.99 (0.90)} & 0.91 (0.86)   & 0.97 (0.90)           \\  
       \hline \hline
     \end{tabular}
     \label{tab:stats}
   \end{table}

To test the quantitative accuracy of the different models, we  also compute the root mean square 
relative deviation, RMSrD, for each (see Table \ref{tab:stats2}).
The results are far from the accuracy of the methods based on the exact solution of the BTE, however some of the 
combinations present interesting results considering their lower computational cost.
$\kappa_l\left(\theta^{(2)}_a,\gamma^{(2)}_a\right)$, $\kappa_l\left(\theta^{(2)}_a,\gamma^{(3)}\right)$ and
$\kappa_l\left(\theta^{(4)}_a,\gamma^{(3)}\right)$ present a RMSrD lower than 100\%.
Moreover, in all cases, the major contribution to this deviation corresponds to the materials with a lower $\kappa_l$
in which, a small absolute error could be transform in a big relative error.

   \begin{table}[h]
     \caption{Root mean square relative deviation (RMSrD) for the material data set. Lowest RMSrD in bold.}
     \centering
     \begin{tabular}[b]{c C{1.2cm} C{1.2cm} C{1.2cm} C{1.2cm} }
       \hline \hline
                & $\theta^{(1)}_a$ & $\theta^{(2)}_a$ & $\theta^{(3)}_a$ & $\theta^{(4)}_a$  \\
       \hline
        $\gamma^{(1)}_a$  &237\% & 120\%                  &  355\% &200\%             \\  
        $\gamma^{(2)}_a$  &230\% &  90\%                  &  435\% &200\%             \\ 
        $\gamma^{(3)}$    &112\% & \textbf{ 75\%} &  269\% & 99\%             \\    
       \hline \hline
     \end{tabular}
     \label{tab:stats2}
   \end{table}

Combining both statistical descriptors, we can conclude that the best results are obtained for 
$\kappa_l\left(\theta^{(2)}_a,\gamma^{(3)}\right)$. 
However, these results should be compared with previous models to evaluate their performance.
To the best of our knowledge, the most established and accurate model to predict $\kappa_l$ using only 
harmonic force constants has been developed by Madsen \textit{ et al.} \cite{Madsen_pssa_2016}. 
They used an approximated value for the phonon scattering time using the Debye 
temperature and the Gr\"{u}neisen parameters.
This method ($\kappa^M_l$) and our combinatorial approach have the same computational cost, the calculation of the harmonic force constants.
The comparison between $\kappa^M_l$ and $\kappa_l\left(\theta^{(2)}_a,\gamma^{(3)}\right)$ is 
depicted in Figure \ref{fig:kappa_panel} (j).
Both models show similar behaviors, although, the statistical descriptors are slightly different. 
$\kappa_l\left(\theta^{(2)}_a,\gamma^{(3)}\right)$ presents a smaller RMSrD than $\kappa^M_l$ (75\% and 89\% respectively) 
and a higher Pearson  correlation (0.99 and 0.98 respectively). 
We can also compare $\kappa_l\left(\theta^{(2)}_a,\gamma^{(3)}\right)$ with less expensive methodologies such  as AGL (see Figure \ref{fig:kappa_panel} (j)).
Although $\kappa^{AGL}_{l}$ captures some of the overall trends and can be used as a fast screening method, this method underestimates $\kappa_l$.
This underestimation can be as large as an order of magnitude which is particularly detrimental to compounds with high  $\kappa_l$ values.

\section{Conclusions}

A high throughput combinatorial method for fast and robust prediction of lattice thermal conductivity is presented. 
Using the QHA-APL implementation \cite{Pinku_sr_2016}, we can compute different definitions for the Debye temperature and the Gr\"{u}neisen parameter. 
These values can be used in a combinatorial approach with the  Slack equation to obtain different values of $\kappa_l(\theta_a,\gamma)$.
We use a set of 42 materials to validate our results and elucidate the best combination of parameters.
Best results are obtained when $\theta^{(2)}_a$ is used, being the most accuarte procedure to predict the acoustic Debye temperature.
$\kappa_l\left(\theta^{(2)}_a,\gamma^{(3)}\right)$ is the most robust and accurate of all the combinations, obtaining qualitatively and 
quantitatively better results than other models in the literature with similar computational cost.
The methodology presented in this paper can be extremely useful because of its quantitative predictive power and its low 
computational cost compared to the calculation of the anharmonic force constants.

\section{Acknowledgments}
We would like to acknowledge support by DOD-ONR (N00014-13-1-0635, N00014-11-1-0136, N00014-09-1-0921) and by DOE (DE-AC02- 05CH11231), specifically the BES program under Grant \# EDCBEE. 
The {\small AFLOW} consortium would like to acknowledge the Duke University - Center for Materials Genomics and the CRAY corporation for computational support.

\newpage
\bibliographystyle{elsarticle-num-names}
\newcommand{\Ozolins}{Ozoli\c{n}\v{s}}

\end{document}


\begin{frontmatter}
  \title{Supplementary Materials: High Throughput combinatorial method for fast and\\ robust prediction of lattice thermal conductivity}
  \author{Pinku Nath$^1$, Jose J. Plata$^1$, Demet Usanmaz$^1$, Cormac Toher$^1$,\\ Marco Fornari$^2$, Marco Buongiorno Nardelli$^3$, Stefano Curtarolo$^{4,\star}$}
  \address{$^{1}$ Department of Mechanical Engineering and Materials Science, Duke University, Durham, North Carolina 27708, USA.}
  \address{$^{2}$ Department of Physics and Science of Advanced Materials Program, Central Michigan University, Mount Pleasant, MI 48858, USA.}
  \address{$^{3}$ Department of Physics and Department of Chemistry, University of North Texas, Denton TX, USA.}
  \address{$^{4}$ Materials Science, Electrical Engineering, Physics and Chemistry, Duke University, Durham NC, 27708, USA.}
  \address{$^{\star}${\bf corresponding:} stefano@duke.edu}
  \begin{abstract}
The lack of computationally inexpensive and accurate ab-initio based methodologies to predict lattice thermal conductivity, without
 computing the anharmonic force constants or time-consuming ab-initio molecular dynamics, is one of the obstacles preventing the accelerated discovery
of new high or low thermal conductivity materials.
The Slack equation is the best alternative to other more expensive methodologies but is highly dependent on two variables: the acoustic Debye temperature, $\theta_a$, and the Gr\"{u}neisen parameter, $\gamma$.
Furthermore, different definitions can be used for these two quantities depending on the model or approximation.
In this article, we present a combinatorial approach to elucidate which definitions of both variables produce the best predictions of the lattice thermal conductivity, $\kappa_l$.
A set of 42 compounds was used to test accuracy and robustness of all possible combinations.
This approach is ideal for obtaining more accurate values than fast screening models based on the Debye model, while being significantly less expensive than methodologies that solve the Boltzmann transport equation. 
 \end{abstract}
  \begin{keyword}
    High-throughput \sep materials genomics \sep Quasi-Harmonic Approximation \sep Lattice thermal conductivity 
  \end{keyword}
\end{frontmatter}

\section{Lattice thermal conductivity}

\begin{landscape}
\begin{tiny}
\begin{longtable}[h]{c r r r r r r r r r r r r r r r}
\caption{Calculated and experimental lattice thermal conductivity for the material data set.}\\
\hline  
\hline  
Formula  &  ICSD & \# SG    & $\kappa_l\left(\theta^{(1)}_a,\gamma^{(1)}\right)$    & $\kappa_l\left(\theta^{(1)}_a,\gamma^{(2)}\right)$  & $\kappa_l\left(\theta^{(1)}_a,\gamma^{(3)}\right)$  & $\kappa_l\left(\theta^{(2)}_a,\gamma^{(1)}\right)$  & $\kappa_l\left(\theta^{(2)}_a,\gamma^{(2)}\right)$ & $\kappa_l\left(\theta^{(2)}_a,\gamma^{(3)}\right)$ & $\kappa_l\left(\theta^{(3)}_a,\gamma^{(1)}\right)$ & $\kappa_l\left(\theta^{(3)}_a,\gamma^{(2)}\right)$ & $\kappa_l\left(\theta^{(3)}_a,\gamma^{(3)}\right)$ & $\kappa_l\left(\theta^{(4)}_a,\gamma^{(1)}\right)$ & $\kappa_l\left(\theta^{(4)}_a,\gamma^{(2)}\right)$ &  $\kappa_l\left(\theta^{(4)}_a,\gamma^{(3)}\right)$  & Exp.\\\hline
\endfirsthead
\hline  
Formula & ICSD &  \# SG   & $\kappa_l(\theta^{(1)}_a,\gamma^{(1)})$    & $\kappa_l(\theta^{(1)}_a,\gamma^{(2)})$  & $\kappa_l(\theta^{(1)}_a,\gamma^{(3)})$  & $\kappa_l(\theta^{(2)}_a,\gamma^{(1)})$  & $\kappa_l(\theta^{(2)}_a,\gamma^{(2)})$ & $\kappa_l(\theta^{(2)}_a,\gamma^{(3)})$ & $\kappa_l(\theta^{(3)}_a,\gamma^{(1)})$ & $\kappa_l(\theta^{(3)}_a,\gamma^{(2)})$ & $\kappa_l(\theta^{(3)}_a,\gamma^{(3)})$ & $\kappa_l(\theta^{(4)}_a,\gamma^{(1)})$ & $\kappa_l(\theta^{(4)}_a,\gamma^{(2)})$ &  $\kappa_l(\theta^{(4)}_a,\gamma^{(3)})$  & Exp.\\ 
\hline  
\endhead
\hline
\endfoot
\hline \hline
\endlastfoot
 \ce{C1}       &     182729   & 227 &    1539.59   &     1521.88    &    1055.56   &     1657.55    &    1345.72    &     933.55    &    1560.49   &     1567.21   &     1086.96    &    1393.34   &      928.36    &     644.45   &     3000 \cite{Morelli_Slack_2006}    \\
 \ce{Ge1}      &     44841    & 227 &      148.57   &      147.46    &      66.70   &       95.52    &      61.47    &      28.19    &     149.19   &      185.53   &       83.72    &     112.05   &       75.98    &      34.70   &       65 \cite{Morelli_Slack_2006}   \\
 \ce{Si1}      &     76268    & 227 &      209.71   &      210.80    &     132.38   &      100.17    &      94.82    &      60.30    &     235.80   &      257.69   &      161.46    &     146.50   &      142.06    &      89.69   &      166 \cite{Morelli_Slack_2006}   \\
 \ce{Ag1Cl1}   &     157535   & 225 &        1.37   &        1.90    &       1.32   &        0.68    &       0.57    &       0.39    &       3.55   &        5.61   &        3.90    &       1.60   &        2.38    &       1.65   &        1  \cite{Maqsood_IJT_2003}  \\
 \ce{Br1K1}    &     52243    & 225 &        1.74   &        1.82    &       1.36   &        1.13    &       1.17    &       0.87    &       3.02   &        4.32   &        3.22    &       1.59   &        1.60    &       1.20   &        3.4 \cite{Morelli_Slack_2006}   \\
 \ce{Br1Na1}   &     44278    & 225 &        5.15   &        5.16    &       2.61   &        2.68    &       2.79    &       1.42    &       6.19   &       13.00   &        6.52    &       4.06   &        4.06    &       2.06   &        2.8 \cite{Morelli_Slack_2006}   \\
 \ce{Br1Rb1}   &     53845    & 225 &        1.42   &        1.53    &       1.03   &        1.28    &       1.21    &       0.81    &       2.36   &        3.63   &        2.43    &       1.40   &        1.48    &       0.99   &        3.8 \cite{Morelli_Slack_2006}   \\
 \ce{Ca1O1}    &     180198   & 225 &       19.79   &       23.16    &      23.84   &       18.10    &      18.38    &      18.88    &      62.81   &       65.16   &       67.59    &      24.36   &       28.05    &      28.93   &       27 \cite{Morelli_Slack_2006}   \\
 \ce{Cl1K1}    &     240522   & 225 &        2.27   &        2.45    &       1.87   &        2.05    &       2.06    &       1.58    &       4.43   &        6.03   &        4.60    &       2.35   &        2.59    &       1.98   &        7.1 \cite{Morelli_Slack_2006}   \\
 \ce{Cl1Na1}   &     240600   & 225 &        3.73   &        4.21    &       2.79   &        3.47    &       3.66    &       2.43    &       6.54   &       10.15   &        6.71    &       3.88   &        4.71    &       3.12   &        7.1 \cite{Morelli_Slack_2006}   \\
 \ce{Cl1Rb1}   &     18016    & 225 &        1.02   &        1.08    &       1.12   &        0.52    &       0.61    &       0.63    &       2.41   &        2.56   &        2.67    &       0.72   &        0.79    &       0.83   &        2.8  \cite{Morelli_Slack_2006} \\
 \ce{F1Li1}    &     53839    & 225 &       11.59   &       14.15    &      13.16   &       10.53    &      11.26    &      10.45    &      36.86   &       42.00   &       39.32    &      10.54   &       11.78    &      10.94   &       17  \cite{Morelli_Slack_2006}  \\
 \ce{F1Na1}    &     52238    & 225 &        9.15   &        9.90    &       7.83   &        8.89    &       9.15    &       7.23    &      19.56   &       25.32   &       20.07    &       9.13   &        9.73    &       7.69   &       18.4 \cite{Morelli_Slack_2006}   \\
 \ce{H1Li1}    &     61751    & 225 &       57.05   &       57.36    &      69.64   &       14.06    &      14.84    &      17.59    &     142.05   &      131.56   &      160.96    &      14.25   &       15.01    &      17.80   &       15  \cite{Morelli_Slack_2006}  \\
 \ce{I1K1}     &     53827    & 225 &        2.39   &        2.39    &       1.20   &        1.03    &       1.03    &       0.52    &       2.76   &        5.71   &        2.84    &       1.67   &        1.67    &       0.84   &        2.6 \cite{Morelli_Slack_2006}   \\
 \ce{I1Na1}    &     52240    & 225 &        3.63   &        3.63    &       1.86   &        1.19    &       1.23    &       0.64    &       4.95   &       10.27   &        5.21    &       2.57   &        2.56    &       1.32   &        1.8 \cite{Morelli_Slack_2006}   \\
 \ce{I1Rb1}    &     53846    & 225 &        1.29   &        1.41    &       0.76   &        1.11    &       1.01    &       0.55    &       1.68   &        3.25   &        1.75    &       1.25   &        1.31    &       0.71   &        2.3 \cite{Morelli_Slack_2006}   \\
 \ce{Mg1O1}    &     159372   & 225 &       36.26   &       40.09    &      40.51   &       32.78    &      34.90    &      35.23    &      87.16   &       93.60   &       95.00    &      42.46   &       48.61    &      49.18   &       60 \cite{Morelli_Slack_2006}   \\
 \ce{O1Sr1}    &     26960    & 225 &       20.44   &       23.66    &      21.93   &        9.43    &       9.97    &       9.18    &      71.54   &       82.56   &       77.00    &      16.95   &       17.72    &      16.39   &       12  \cite{Morelli_Slack_2006}  \\
 \ce{Be2C1}    &     616184   & 225 &      108.44   &      107.75    &      98.51   &      103.11    &     102.05    &      93.28    &     165.04   &      175.42   &      160.45    &     103.54   &      102.56    &      93.75   &       23.43 \cite{Thermtest}    \\
 \ce{Ca1F2}    &     40938    & 225 &        6.62   &        7.03    &       8.26   &        4.66    &       4.67    &       5.46    &      17.76   &       17.52   &       20.82    &       7.65   &        8.37    &       9.86   &        8.5 \cite{Palchoudhuri_ssc_1989}   \\
 \ce{Al1As1}   &     606008   & 216 &      164.70   &      164.82    &      86.23   &       52.91    &      44.06    &      23.86    &     249.31   &      249.49   &      129.77    &      85.08   &       85.05    &      45.10   &       98 \cite{Morelli_Slack_2006}   \\
 \ce{Al1P1}    &     609019   & 216 &      180.93   &      179.34    &      92.70   &       95.96    &      86.51    &      45.33    &     244.49   &      266.81   &      137.21    &     150.15   &      141.59    &      73.46   &       90 \cite{Spitzer_JPCS_1970}   \\
 \ce{Al1Sb1}   &     609290   & 216 &       95.73   &       96.10    &      74.20   &       14.29    &      14.37    &      11.48    &     149.40   &      149.98   &      115.40    &      31.79   &       31.92    &      25.01   &       56 \cite{Morelli_Slack_2006}   \\
 \ce{As1Ga1}   &     53964    & 216 &      114.56   &      111.09    &      48.18   &       75.67    &      50.32    &      22.12    &     130.39   &      135.94   &       58.82    &      96.11   &       72.09    &      31.46   &       45 \cite{Morelli_Slack_2006}   \\
 \ce{As1In1}   &     165462   & 216 &       54.99   &       55.15    &      31.18   &       21.57    &      18.98    &      10.98    &      68.99   &       69.25   &       39.03    &      34.96   &       32.27    &      18.42   &       30 \cite{Morelli_Slack_2006}   \\
 \ce{Br1Cu1}   &     30090    & 216 &        8.31   &        8.03    &       3.11   &        4.26    &       3.46    &       1.36    &      10.36   &       10.72   &        4.13    &       9.92   &       11.99    &       4.61   &        1.25 \cite{CRHamond_CRC_2005}    \\
 \ce{Cd1S1}    &     290009   & 216 &       43.93   &       44.19    &      32.27   &        6.31    &       6.88    &       5.23    &      69.57   &       73.11   &       53.13    &      20.02   &       20.14    &      14.89   &       16  \cite{Morelli_Slack_2006}  \\
\ce{Cl1Cu1}   &     78270     & 216 &        6.10   &        6.30    &       4.46   &        1.16    &       1.29    &       0.94    &       8.51   &       12.08   &        8.52    &       5.99   &        6.14    &       4.35   &        0.84 \cite{CRHamond_CRC_2005}   \\
 \ce{Cu1I1}    &     163427   & 216 &       14.63   &       12.52    &       3.03   &       10.85    &       6.89    &       1.69    &      15.62   &       14.94   &        3.61    &      11.40   &        7.35    &       1.80   &        1.68 \cite{CRHamond_CRC_2005}    \\
 \ce{Fe1Sb1V1} &     53539    & 216 &       43.83   &       43.84    &      28.72   &       22.43    &      22.06    &      14.55    &      68.44   &       84.85   &       55.38    &       0.00   &        0.00    &       0.00   &       13  \cite{Young_jap_2000}  \\
 \ce{Ga1P1}    &     77088    & 216 &      266.36   &      265.99    &     113.90   &      125.13    &      81.46    &      36.09    &     351.88   &      351.39   &      149.76    &     152.78   &      146.69    &      63.67   &      100 \cite{Morelli_Slack_2006}   \\
 \ce{Ga1Sb1}   &     41675    & 216 &       58.07   &       58.07    &      33.44   &       25.23    &      19.75    &      11.62    &      73.89   &       73.89   &       42.42    &      37.47   &       32.45    &      18.86   &       40 \cite{Morelli_Slack_2006}   \\
 \ce{Hg1Se1}   &     31087    & 216 &        2.47   &        2.49    &       3.52   &        0.31    &       0.42    &       0.56    &       4.33   &        4.16   &        5.92    &       1.38   &        1.39    &       1.94   &        3 \cite{Whitsett_PRB_1973}   \\
 \ce{In1P1}    &     165466   & 216 &      179.87   &      179.84    &      91.75   &       40.34    &      30.80    &      16.83    &     254.70   &      254.65   &      129.11    &      69.31   &       69.10    &      36.23   &       93 \cite{Morelli_Slack_2006}   \\
 \ce{O1Zn1}    &     647683   & 216 &      329.31   &      330.57    &     131.69   &       90.82    &      72.16    &      30.38    &     273.47   &      624.17   &      246.03    &     156.96   &      157.57    &      64.03   &       60 \cite{Morelli_Slack_2006}   \\
 \ce{S1Zn1}    &     108733   & 216 &      149.65   &      149.44    &      52.06   &       71.53    &      49.10    &      17.68    &     219.42   &      219.10   &       75.82    &     106.54   &      105.99    &      37.21   &       27 \cite{Morelli_Slack_2006}   \\
 \ce{Se1Zn1}   &     181761   & 216 &       70.42   &       65.33    &      22.36   &       43.72    &      33.21    &      11.53    &      88.99   &       98.99   &       33.70    &       0.00   &        0.00    &       0.00   &       19 \cite{Morelli_Slack_2006}   \\
 \ce{Te1Zn1}   &     184485   & 216 &       45.82   &       45.71    &      16.16   &       26.95    &      19.30    &       6.96    &      68.78   &       68.75   &       24.16    &      38.13   &       34.27    &      12.18   &       18  \cite{Morelli_Slack_2006}  \\
 \ce{C1Si1}    &     618777   & 186 &    1016.65   &     1300.76    &     318.00   &      571.85    &     600.12    &     151.94    &    1207.34   &     2081.67   &      501.90    &     679.22   &      744.18    &     186.23   &      360 \cite{Ioffe_Inst_DBp}   \\
\end{longtable}
\end{tiny}
\end{landscape}

\newpage

\bibliographystyle{elsarticle-num-names}
\newcommand{\Ozolins}{Ozoli\c{n}\v{s}}